\documentclass[manuscript]{aastex} 
\usepackage{emulateapj5}  
\newcommand{\smmsrcf}{\mbox{SMM J02399$-$0136}}
\newcommand{\smmsrc}{\mbox{CX1}}
\newcommand{\smmsrcp}{\mbox{02399$-$0136}}
\newcommand{\smmsecf}{\mbox{SMM J02400$-$0134}}
\newcommand{\smmsecp}{\mbox{02400$-$0134}}

\newcommand{\lrgsrc}{\mbox{LRG J0239$-$0134}}
\newcommand{\lrgsmmf}{\mbox{SMM J02399$-$0134}}
\newcommand{\lrgsmmp}{\mbox{02399$-$0134}}
\newcommand{\lrgsmm}{\mbox{CX2}}

\newcommand{\Msun} {\mbox{M$_{\odot}$}}
\newcommand{\Lsun} {\mbox{L$_{\odot}$}}

\begin{document}
\slugcomment{Accepted 2000 July 27 for publication in The Astrophysical Journal Letters}
\title{Detection of  X-ray Emission from Gravitationally Lensed Submillimeter Sources in the Field of Abell 370}
\author{ M. W. Bautz,\altaffilmark{1} M. R. Malm,\altaffilmark{1}
F. K. Baganoff,\altaffilmark{1} G. R. Ricker,\altaffilmark{1}
C. R. Canizares,\altaffilmark{1} \\ W. N. Brandt,\altaffilmark{2} 
A. E. Hornschemeier,\altaffilmark{2} and G. P. Garmire\altaffilmark{2}}

\altaffiltext{1}{Center for Space Research, Massachusetts Institute of Technology, Cambridge, MA 02139}
\altaffiltext{2}{Department of Astronomy \& Astrophysics, 525 Davey Laboratory, The Pennsylvania State University, University Park, PA 16802}
\begin{abstract}
We report the detection by Chandra of \lrgsmmf\, and \smmsrcf, two distant
($z=1.06$ and $z=2.81$, respectively) submillimeter sources gravitationally magnified by the galaxy cluster 
Abell 370. These are high-significance ($> 7\sigma$) X-ray detections of the
high-redshift submillimeter source population. 
The X-ray positions are 
coincident with the optical positions to within one arcsecond.
The X-ray spectra, while of low signal-to-noise ratio, are quite hard.
Absorbed power law models with fixed photon indices of $\Gamma=2.0$
imply local absorbing columns $>2 \times 10^{23}$ cm$^{-2}$ and
unabsorbed luminosities $>10^{44}$ erg s$^{-1}$  in both sources.
These results imply that nuclear activity
is responsible for the bulk of the luminosity  in $\lrgsmmf$, 
and for at least 20\% of the luminosity of \smmsrcf, consistent
with previous optical observations.
We also place an upper limit on the X-ray flux of a third submillimeter source,
\smmsecf. Considered together with previously published  Chandra upper 
limits on 
X-ray flux from submillimeter sources, our results imply that $20^{+30}_{-16}$ \%
of submillimeter sources exhibit X-ray emission from  AGN (90\% confidence), 
consistent with expectations of their contribution to the diffuse X-ray background.
\end{abstract}
\section{Introduction}
Deep surveys with the Submillimetre Common User Bolometer Array
(SCUBA: Holland et al. 1998) in the fields of rich clusters~\citep{sm98} 
and in  blank fields (e.g., Hughes et al. 1998; Barger, Cowie \& Sanders 1999; Eales et al. 1999) have 
revealed a large population of  luminous, star-forming galaxies at 
high-redshift.  This population produces a substantial fraction of the 
diffuse infrared background \citep{fx98}, and, while  
the contribution of AGN to its total
luminosity is uncertain, accounts for at least as much star formation 
as has been inferred from UV/optical counts \citep{dw98,bl99}.
Besides their significance for the star-formation history of the 
universe, a subset of the submillimeter (sub-mm) source population  
may also host the obscured AGN thought to be responsible for the diffuse 
X-ray background (Almaini, Lawrence \& Boyle 1999, hereafter ALB99; Hasinger 2000; Fabian et al. 2000). 
X-ray background models (ALB99; Gunn \& Shanks 1999) and optical
spectroscopy of sub-mm sources (Barger et al. 1999, hereafter B99) both 
suggest that $\sim 5$\% -- $20$\% of SCUBA sources should contain an AGN.

Deep X-ray observations with the Chandra X-ray Observatory
can test this connection directly.  
We report here high-significance 
X-ray detections  of bright sub-mm sources at high redshift, and
the first X-ray spectral constraints on this population.
The three sub-mm sources we discuss,
\smmsrcf, \lrgsmmf, and \smmsecf, are gravitationally magnified by 
Abell 370 and are detected at greater than  $4\sigma$ significance by 
SCUBA (Smail et al. 1997; B99). We detect the first two of these, 
hereafter designated \smmsrc\ and \lrgsmm, respectively. 
Both X-ray-detected sources were
previously identified with high-redshift 
($z=2.81$ and $z=1.06$, respectively) star-forming galaxies containing AGN
(Ivison et al., 1998, hereafter I98; Soucail et al., 1999, hereafter S99). 
We place an upper limit on the X-ray flux from the third, which 
has no optical counterpart to a limiting  magnitude of  
$I \sim 26$ (Smail et al. 1998; B99).

\section{Observations}
Chandra observed Abell 370 with the ACIS S3 detector ~\citep{ga00} 
in 1999 October for 93 ks. The net useful exposure time excluding
periods of high background and bad aspect is 66.6 ks.
Events in  ASCA grades 1, 5, and 7 were excluded, and 
unless otherwise noted, our analysis is restricted to the 
0.3--7 keV spectral range.

{\it Source Detection and Astrometry: }
We ran the Chandra X-ray Center's {\tt wavdetect} wavelet source detection 
program on an $8.4^{\prime}\times8.4^{\prime}$ field containing the cluster. 
We used wavelet kernel 
scales of 1--8 arcsec, and set  the threshold detection significance parameter to  10$^{-7}$.  
Two of the 34 sources detected
are positionally coincident with  sub-mm sources described above 
to within 2.6 arcsec. 

Table~\ref{tab:srcpos} lists the X-ray positions and  compares them with the best 
available optical  positions (source L1 of I98 for \smmsrc ; S99 for \lrgsmm ) 
for the identified sub-mm sources. 
Formal random errors in the X-ray positions are of order 0.2$^{\prime\prime}$ in each
axis.  
Systematic errors of  $\sim1.5^{\prime\prime}$ may affect the Chandra aspect solution currently available for this observation.  We compared the 
positions of other compact sources detected in the Chandra image to 
various optical catalogs and found
4 stellar objects in the APM catalog with positions within 3$^{\prime \prime}$
of sources detected in the ACIS S3 field. The mean
coordinate differences between the Chandra
and APM frames derived from these objects are listed in Table~\ref{tab:srcpos}.
The standard deviation of each of these mean differences is 
0.5$^{\prime \prime}$, and the external accuracy of the APM astrometry is claimed to be $\sim 0.5^{\prime \prime}$ ~\citep{apm}.

\begin{figure*}
\figurenum{1}
\epsscale{0.75}
\label{fig:xopt}
\plotone{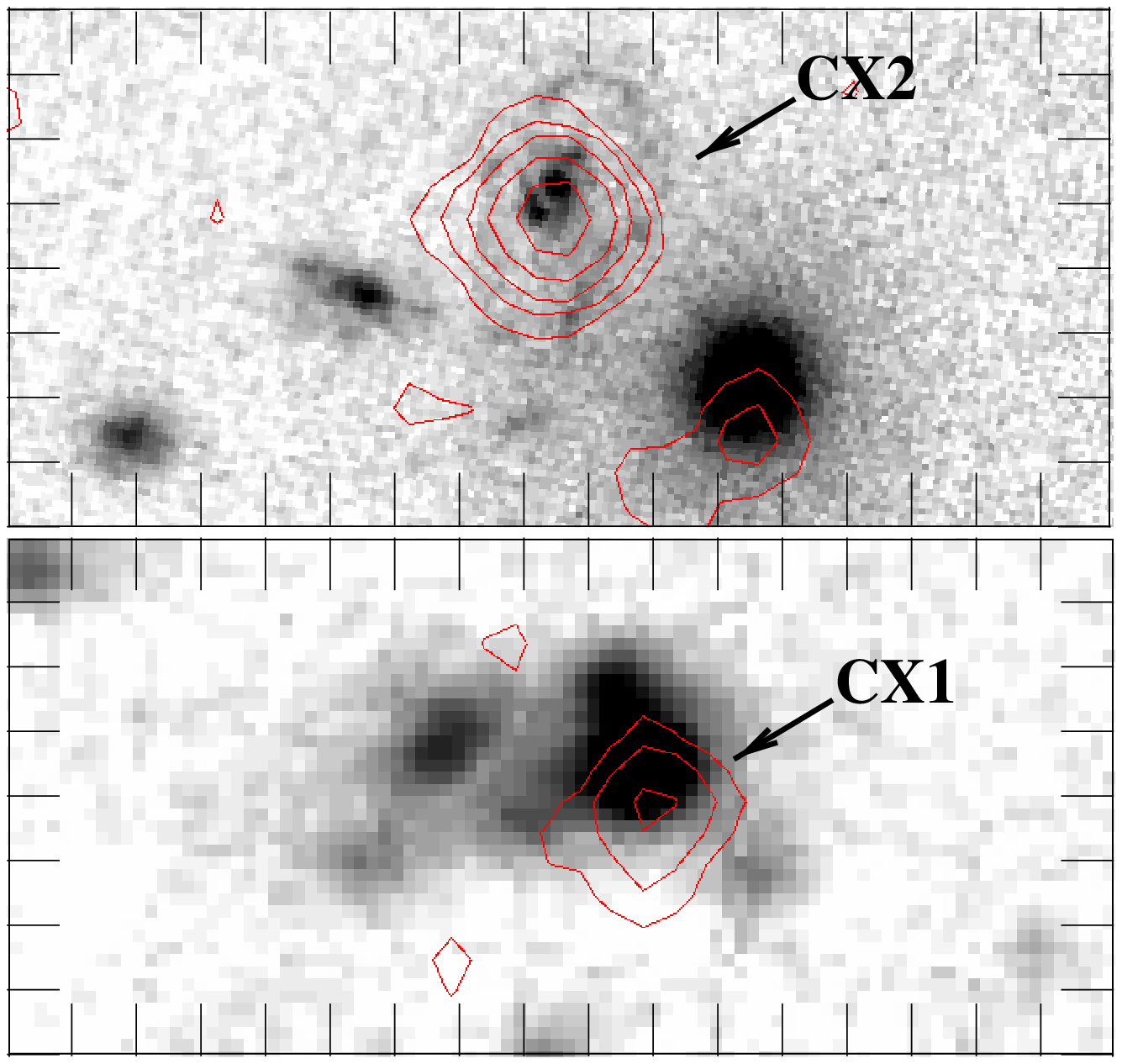}
\caption{Smoothed Chandra X-ray surface brightness contours superposed on 
optical images of submillimeter
sources. {\it Top:} Hubble Space Telescope F675W image of the ring galaxy \lrgsrc\ identified
with the \lrgsmmf\ (\lrgsmm; S99). {\it Bottom:}
Keck R-band image of the optical counterpart (L1 of I98) 
of \smmsrcf\ (\smmsrc; Barger \& Cowie, 2000). 
North is at the top, East on the left. The tic marks are  1$^{\prime\prime}$ 
apart.
The lowest contour level is twice the local cluster background, and
the contours are logarithmically spaced by a factor of 2.
In each case the submillimeter source has been identified with 
the brightest optical object within the labelled X-ray 
contours.}
\end{figure*}

We adopt the mean Chandra$-$APM offset as the Chandra boresight error, 
and find that the Chandra and  optical positions for both sources
agree within 1$^{\prime\prime}$.  X-ray source positions (from other ACIS 
detectors) for four objects in the USNO catalog, so corrected, have a mean 
radial error  $< 1.1^{\prime\prime}$. We note that 
the 3 mm radio position of \smmsrc\ ~\citep{fr98} agrees with both the optical and X-ray positions to within 1$^{\prime\prime}$.
We note also that 14 X-ray sources at least as 
bright as \smmsrc\ are detected in the Abell 370 field; the
probability of any of these fortuitously lying within 1.5 arcec of one of the
sub-mm sources is less than $4 \times 10^{-4}$. 
X-ray surface brightness contours from the gaussian-smoothed 
($\sigma=0.5^{\prime\prime}$), boresight-corrected 
Chandra image are overlayed on optical images in Figure~\ref{fig:xopt}.

{\it Photometry and Spectroscopy:}
Net source counts in $5^{\prime\prime} \times 5^{\prime\prime}$ apertures
for detections, and 99\% confidence upper limits in 2$^{\prime\prime}$
radius apertures, with corresponding background levels, are listed
in Table~\ref{tab:srcpos}.
Although the local  cluster emission exceeds the particle  background
by a factor of $4- 5$ at the source locations, the two
(hard-band) detections are highly significant ( $> 7\sigma$ equivalent).
Upper limits were computed using the prescription of~\citet{kbn91}. 
The celestial coordinates of \smmsecf\ are taken to be
$\alpha = 2^{h} 39^{m} 57.88^{s}$, $\delta = -01^{\circ} 
34^{\prime} 45.1^{\prime \prime}$ (J2000), 
derived from the finding chart of B99.
X-ray fluxes are derived 
from the hard-band counts assuming a power law spectrum with a photon index of $\Gamma=0$. The 850$\mu$m  flux densities are from B99.

In spite of the small count totals, 
we can constrain the spectra of the detected sources. 
Both sources are quite hard, as  neither is detected above the
cluster emission in the  0.3--1.5 keV  band by {\tt wavdetect}. 
We fit a series of  simple {\tt XSPEC} \citep{ar96} models
to pulse-height spectra extracted from 
$5^{\prime\prime} \times 5^{\prime\prime}$ square apertures. We  modelled the 
background as the sum of i) a power law component and fluorescent lines, 
accounting for the particle background; ii) a cool thermal
component, representing  Galactic emission; and iii) a hot thermal component
representing the cluster. The relative amplitudes and shapes of the
various background components were determined from suitable 
regions of the cluster image,
and  normalizations were determined locally for 
each source. The resulting fixed background model was added to the source model and 
the maximum likelihood statistic was minimized to identify the best-fit source parameters.

Optical and infrared data 
suggest that these objects contain AGN (see Section~\ref{sec:disc}), 
so we fit power law models with absorbtion by the Galaxy
(fixed at N$_{H} = 2.5 \times 10^{20}$ cm$^{-2}$; Stark et al. 1992) and
by neutral material at the source. If the source absorbing column 
($N_{H,s}$) is fixed at zero, the 
best-fit power law index ($\Gamma$) is consistent
with zero, within wide limits,  for both sources.
In the case of \lrgsmm\ this model systematically over-predicts the data at 
both low and high energies, suggesting that additional absorption is required.
The fit is better if both $N_{H,s}$ and $\Gamma$ are varied, 
but $\Gamma$ is very poorly constrained. In the case of \smmsrc,
an $F$-test shows that the addition of local absorption does not produce a 
statistically significant improvement in the fit. For both sources, the
indeterminancy in
$\Gamma$ is accompanied by a corresponding uncertainty in the absorbing
column  and in the intrinsic luminosity of the source. 

Faced with this ambiguity, we fix the photon number index in our fits.
The X-ray spectra of low-redshift AGN generally exhibit 
(unabsorbed) continuum slopes in a restricted range ($ 1.5 \le \Gamma
\le 2.5$; e.g., Nandra et al. 1997; Turner et al. 1997). Here we consider models with $1.5 \le \Gamma \le 2.0$. 
Any reflected component present is likely to be much harder, and, if 
the obscuring column is sufficiently high, 
will dominate the observed emission, at least at lower energies.  

Results of fits are presented 
in  Table~\ref{tab:spec}, where values for fixed $\Gamma=1.5$ 
and $\Gamma=2.0$ are separated by slashes, 
and the 90\% confidence envelope for these two cases is given in parentheses.
To allow interpretation of the X-ray flux
either as directly transmitted (through an absorbing column)  or 
as reflected emission, we quote both  ``unabsorbed'' and ``observed''
2--10 keV luminosity in the source
rest frame.  All tabulated luminosities  have been 
corrected for the estimated gravitational magnification as discussed in 
Section~\ref{sec:disc} below.

\section{Discussion}
\label{sec:disc}
From the  positional coincidences discussed above we may reasonably 
identify the Chandra and sub-mm sources.
Here we discuss the role of the AGN 
in the bolometric luminosity of these sources, and  consider the 
implications of our results for the expected connection between the 
X-ray and sub-mm backgrounds.

\subsection{The AGN Contribution to Bolometric Luminosity}
The role of AGN in the energetics of ultraluminous infrared galaxies remains 
an open question, particularly
in the case of the objects thought to comprise the high-redshift 
sub-mm source population. In the following discussion, we adopt
a gravitational magnification of 2.5 for both sources. The uncertainty in
this  value is no more than a factor of 2 in the case of \smmsrc\ and
about  $\pm$10\% in the case of \lrgsmm\,(Kneib et al, 1993; I98; S99).

{\em \smmsrcf\,(\smmsrc): }
I98 presented comprehensive optical and infrared data which determined 
the redshift and showed emission from a narrow-line, dust-obscured AGN.
They report detection by ISO at 15 $\mu$m but were
unable to determine the relative importance of star formation and the AGN
in producing the enormous bolometric luminosity of this object 
(L$_{FIR} \sim 5 \times 10^{12}$ \Lsun , 
referred to our cosmology.)  Frayer et al. (1998) detected CO line emission 
in \smmsrc,  implying a  large mass of molecular 
gas (10$^{10}$ \Msun) and a relatively
small ratio of far-infrared (FIR) to CO luminosity, indicating
that star formation is important in this source. They conclude 
from the relatively low FIR-to-4.85 GHz flux ratio that 50\% $\pm$ 25\% of
the infrared luminosity may be powered by the AGN. 

We find that the ratio of X-ray to bolometric 
luminosity for this source exceeds 1.5\% for $\Gamma >1.5$. Although this limit
is rather low for radio-quiet quasars~\citep{el94}, we note that at 
least two luminous type 2 AGN,
IRAS 23060+0505~\citep{br97} and IRAS 20460+1925~\citep{og97}, have 
comparable or lower 
values of  L$_{X}$/L$_{bol}$.  Adopting the median bolometric correction 
L$_{X}$(1-10 keV)/L$_{bol} =  0.05$ from~\citet{el94}, we find 
L$_{AGN} \simeq 9 \times 10^{45}$ erg s$^{-1}$, about 40\% of the 
FIR luminosity.  Between 20\% and 80\% of the luminosity of \smmsrc\ is 
attributable to the obscured AGN if $1.5 \le \Gamma \le 2.0$.

The large column density ( $N_{H,s} > 10^{24}$ cm$^{-2}$
if $\Gamma = 2.0$)  inferred for \smmsrc\ suggests that the obscuring
material may be Compton-thick, in which case we must
interpret the  X-ray emission as predominantly reflected (rather than obscured)
radiation from the 
central source. From the ``observed'' luminosity in Table~\ref{tab:spec},
it follows that if the
X-ray emission is due to reflection from cold gas with an albedo of 0.022
\citep{iw97}, then the nucleus  in \smmsrc\ is of quasar luminosity.

\begin{figure*}
\figurenum{2}
\epsscale{1.0}
\plotone{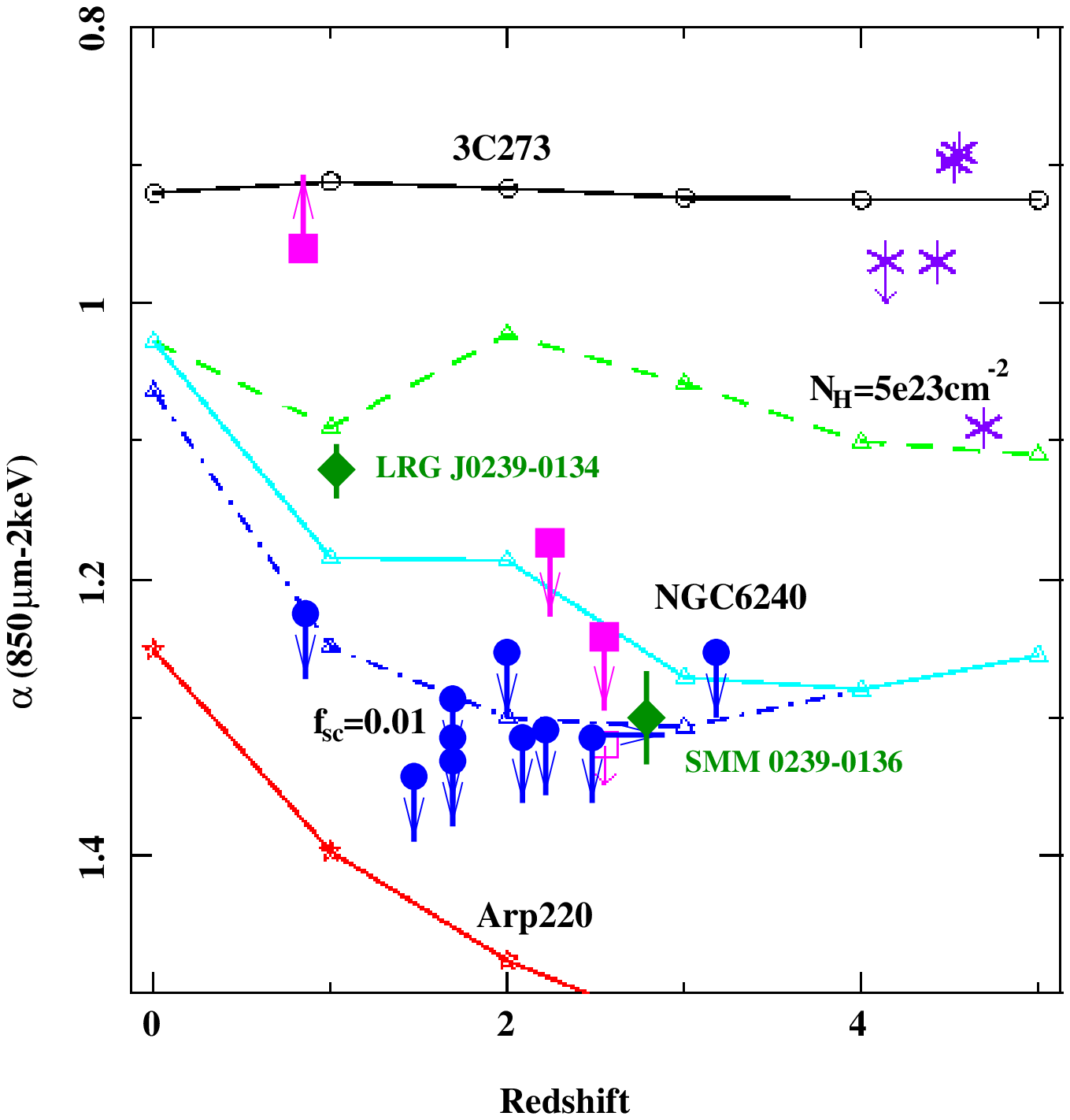}
\caption{ Expected submillimeter-to-X-ray flux ratio as a function of 
redshift for the quasar 3C273, the absorbed, IR-luminous galaxy NGC6240, 
the starburst galaxy Arp 220 (curves), and five high-redshift quasars ($\ast$);
limits from Chandra (squares: Fabian
et al., 2000; circles: Hornschemeier et al., 2000, both modified as
described in the text), and Chandra detections reported here 
(diamonds with error bars). Alternative curves for NGC6240 with less
internal absorption ($N_{H} = 5 \times 10^{23}$) or a smaller scattered
flux fraction (f$_{sc} = 0.01$) are also shown.
Adapted from Fabian et al. (2000) and Hornschemeier et al. (2000). }
\label{fig:andy}
\end{figure*}

We can independently estimate the importance of the AGN in \smmsrc\ 
by comparing its sub-mm-to-X-ray flux ratio with that of the 
ultraluminous infrared galaxy NGC 6240,
in which a highly-obscured,  powerful AGN dominates the bolometric 
luminosity~\citep{vit99}.
Following Fabian et al. (2000), we characterize this ratio
as an equivalent energy index $\alpha$, defined so that spectral flux
density $F_{\nu} \propto \nu^{-\alpha}$. Taking $F_{\nu}$ at 850 $\mu$m and
2 keV from Table~\ref{tab:srcpos},
we find $\alpha = 1.30 \pm 0.03$. This value is
less than that expected of  NGC6240 (observed at $z=2.8$) 
by about a factor of 2 
(corresponding to $\Delta\alpha \sim 0.05$); see 
Figure~\ref{fig:andy}. This might mean either that the fraction of scattered
radiation is lower, or that the AGN is relatively less luminous, 
in \smmsrc\ than in NGC 6240.

{\em \lrgsmmf\ (\lrgsmm): }
S99 identified the ring galaxy \lrgsrc\ with \lrgsmmf, measured its redshift, 
reconstructed its intrinsic shape, and noted that the ring feature is
indicative of interaction-induced star formation.
The optical spectrum and the ISO mid-IR colors led S99 to
identify the source as a Seyfert type 1.
From a Keck LRIS spectrum, B99 classify it as a Seyfert 1.5.

The X-ray luminosity and spectrum of this source confirm the presence
of a powerful, obscured AGN. Applying a bolometric correction 
to the unabsorbed X-ray luminosity yields $L_{bol} \sim 3 \times 10^{45}$ erg s$^{-1}$ which dominates the luminosity in the 
sub-mm (B99) and the mid-IR~\citep{so99}.
For \lrgsmm, $\alpha$  = 1.12, which is quite comparable to the value
expected if NGC6240  were observed at $z=1.06$, allowing for the relatively
smaller observed absorbing column in \lrgsmm\ \citep{fa00};
see Figure~\ref{fig:andy}.

\subsection{ The X-ray/Submillimeter Connection}

Two other searches with Chandra for X-ray emission from sub-mm
sources have yielded only upper limits or detections marginal in 
the sub-mm or the X-ray bands.
Fabian et al. (2000) report upper limits for six SCUBA sources
in two cluster fields, with typical flux upper limits (uncorrected for gravitational magnification) in the deeper field of
4--5 $\times 10^{-15}$ erg s$^{-1}$ cm$^{-2}$ (2--7 keV). Hornschemeier et al.
(2000) report upper limits for 10 SCUBA sources from a very deep (166 ks)
Chandra image containing the Hubble Deep Field (HDF), with typical 
upper limits  0.8 $\times 10^{-15}$ erg s$^{-1}$ cm$^{-2}$
(2--8 keV). Given the cluster X-ray emission, we estimate 
our 99\% confidence detection threshold
for sources near the center of A370 to be about 1.5 $\times 10^{-15}$
 erg s$^{-1}$ cm$^{-2}$ in the 1.5--7 keV band for sources with the spectrum
($\Gamma=2.0$, and no intrinsic absorption) assumed by both groups.

We must estimate the fraction of sub-mm sources that have detectable
AGN using a sample  that has been observed
to an explicit sub-mm-to-X-ray flux ratio (i.e., to a particular
value of $\alpha$, as defined above). We adopt 
our lower limit  for \smmsecf, viz. $\alpha > 1.29$,
(as determined from the hard band flux). While none of the limits
on $\alpha$ from Fabian et al., and only one of
the limits from Hornschemeier et al. is this high, we note 
that both groups compute flux 
assuming $\Gamma=2.0$ {\em and no absorption at the source}.
We observe spectra much harder than this assumed model. 
Accordingly, we recompute the previously published flux limits
using a flat ($\Gamma = 0$, unabsorbed) spectrum, and 
find that the (hardband) X-ray flux density limits drop by about a 
factor of two. After making this adjustment, we find 
that 7 of the HDF limits are sufficiently stringent to have detected a source
with  $\alpha < 1.29$.  
In this sample, then, 2 of 10 sub-mm sources ($20^{+30}_{-16}$\%  
where  errors give the  central 90 percent confidence interval for a binomial
distribution) have detectable X-ray emission.
Fisher's exact probability test~\citep{pe99} shows that 
the probability of finding the observed detection rates, 
given the identical detection efficiency 
in Abell 370 and the HDF, is modest (6.7\%) but not implausible.

After this paper was submitted, Severgnini et. al. (2000) reported 
that 1 of 9 sources detected with SCUBA in the Hawaii SA13 field by
\citet{bar99} was also detected in the Chandra survey of SA13 by 
Mushotzky et al. (2000). The detected source has $\alpha = 1.06$, 
based on the hard-band flux, and the lower limits on $\alpha$ for the other
sources span $ 1.1 < \alpha < 1.2$. Thus the SA13 data are not sensitive
to as low an X-ray-to-sub-mm flux ratio as are ours. If we nevertheless
pool these sources, together with all other Chandra observations
of sub-mm sources discussed above, without regard to sensitivity and 
including marginal detections, 
we find that 7 of 32 sub-mm sources have Chandra counterparts.

The X-ray-detected fractions in both samples are quite consistent with
the predictions of ALB99, 
who model the IR emission of the AGN which produce the diffuse
X-ray background  and find  that 10\% -- 20\% of sources detected at 850$\mu$m
should be AGN. Similarly, the ``conservative'' models of Gunn and 
Shanks (1999) predict an AGN fraction of  5\% -- 15\%, which is 
also consistent with our results.

In summary, we have detected powerful, hard X-ray sources in 
two luminous, high-redshift sub-mm sources.
Both of these objects were previously known, from optical
spectra, to contain AGN, and in both the AGN  are probably responsible for
a substantial fraction of the bolometric luminosity, though
star formation is clearly important in \smmsrcf.  The
proportion of sub-mm sources detected by Chandra to date is 
consistent with models which synthesize the cosmic X-ray 
background from obscured AGN at high redshift.

We thank R. Mushotzky for many helpful 
discussions, and L. Cowie and A. Barger for  the Keck image of \smmsrc. 
This work was supported by NASA under contracts 
NAS-8-37716, NAS-8-38252 and 1797-MIT-NA-A-38252. WNB acknowledges the
support of NSF Career grant AST-9983783.

\begin{table*}
\caption{Observed Properties of Submillimeter Sources \label{tab:srcpos}}
\begin{small}
\begin{tabular} { l c  c c c c c c}
\tableline \tableline
SMM ID\tablenotemark{a} & CXO Name  & 
$\delta$RA\tablenotemark{b} & $\delta$Dec\tablenotemark{b} & 
\multicolumn{2}{c}{Counts\tablenotemark{c}} & F$_{2keV}$\tablenotemark{d} &
S$_{850}$\tablenotemark{e} \\
(J2000)  & (J2000) & ($^{\prime\prime}$)          & ($^{\prime\prime}$)      & 
  1.5--7keV & 0.3--1.5keV &  & \\
\tableline
\smmsrcp\,(\smmsrc) & 023951.9$-$013558 &  0.57 & $-0.61$ & 
 25.1/5.9 & $<17$/4.5   & 0.7 (0.4,0.9) & 25.3 \\
\lrgsmmp\,(\lrgsmm) & 023956.6$-$013426 & 2.55 & $-0.44$ & 
 115/8.1  & $<19.6$/5.8 &  3.2 (2.7,3.7)& 11.0\\
\smmsecp & & & &  $<$ 9.5/2.3 & $<$ 7.2/3.9 &  $<0.26$ & 7.6\\
\multicolumn{2}{l}{Chandra {\it vs.} APM Boresight Difference} & 1.6 & $-0.5$ &  &  & \\
\tableline
\end{tabular}
\end{small}
\tablecomments{ 
$^{a}$Submillimeter Source Name (designation in text); 
$^{b}$Chandra coordinate - optical coordinate; 
$^{c}$Source Counts or Upper Limits/Background in 66.6 ksec,  
99\% confidence upper limits in $2^{\prime\prime}$ radius aperture; 
$^{d}$Flux density at 2 keV,  $10^{-15}$ erg s$^{-1}$ cm$^{-2}$ keV$^{-1}$, 
90\% confidence interval for detections; 99\% confidence upper limit;
$^{e}$Flux density at 850$\mu$m, mJy.}

\end{table*}

\begin{table*}
\caption{Spectral Model Parameters \label{tab:spec}}
\begin{small}
\begin{tabular} { l c c c }
\tableline \tableline
Source  & N$_{H,s}$\tablenotemark{a} & L$_{X,observed}$\tablenotemark{b} &  L$_{X,unabsorbed}$\tablenotemark{b} \\
\tableline
\smmsrc  & 9/13 (5,23) &0.31/0.33 (0.12,0.62)  &3.0/6.4 (1.7,8.8)  \\
\lrgsmm  & 2.1/2.7 (1.6,3.5) &  0.43/0.45 (0.40,0.56)  & 1.0/1.4 (0.80,1.8) \\
\tableline 
\end{tabular}
\tablecomments{
$^{a}$Equivalent hydgrogen column density at source, $10^{23}$cm$^{-2}$;
$^{b}10^{44}$ h$_{65}^{-2}$ erg s$^{-1}$, 2-10 keV source frame,
corrected for gravitational magnification of 2.5.
}
\end{small}
\end{table*}

\begin{thebibliography}{99}
\bibitem[Almaini, Lawrence \& Boyle(1999)]{al99} Almaini, O., Lawrence, A. X.
\& Boyle, B. J., 1999, \mnras, 305, L59 (ALB99)
\bibitem[Arnaud(1996)]{ar96} Arnaud, K., 1996, Jacoby, G. \& Barnes, J., eds., Astronomical Data Analysis Software and Systems V; ASP Conference Series, 101,17
\bibitem[APM(1998)]{apm} APM Website, {\tt http://www.ast.cam.ac.uk/\verb+~+apmcat}
\bibitem[Barger, Cowie \& Sanders(1999)]{bar99} Barger, A. J., Cowie, L.L. \& Sanders D. B., 1999, \apjl, 518, L5
\bibitem[Barger et al.(1999)]{bar99b} Barger, A. J., Cowie, L.L, Smail, I., Ivison, R.J., Blain, A. W., \& Kneib, J.-P., 1999, \aj, 117, 2656 (B99)
\bibitem[Blain et al.(1999)]{bl99} Blain, A. W., Smail, I., Ivison, R. J., and Kneib, J.-P., 1999, \mnras, 302, 632
\bibitem[Brandt et al.(1997)]{br97} Brandt, W.N., Fabian, A.C., Takahashi, K.,
Fujimoto, R., Yamashita, A., Inoue, H. \& Ogasaka, Y., 1997, \mnras, 290, 617
\bibitem[Cowie \& Barger(2000)]{co00} Cowie, L. \& Barger, A., 2000, private communication
\bibitem[Dwek et al.(1998)]{dw98} Dwek, E., et al., 1998, \apj, 508, 106
\bibitem[Eales et al.(1999)]{ea99} Eales, S., Lilly, S., Gear, W., Dunne, L.,
Bond, J., Hammer, F., LeFevre, O. \& Crampton, D., 1999,\apj, 515, 518
\bibitem[Elvis et al.(1994)]{el94} Elvis, M., Wilkes, B. J., McDowell, J. C., 
Green, R. F., Bechtold, J., Willner, S.P., Oey, M.S., Polomski, E., \& Cutri, R., 1994, \apjs, 95, 1
\bibitem[Fabian et al.(2000)]{fa00} Fabian, A. C., et al. 2000, \mnras, 
(submitted)
\bibitem[Fixsen et al.(1998)]{fx98} Fixsen, D.J. et al., 1998, \apj, 508, 123
\bibitem[Frayer et al.(1998)]{fr98} Frayer, D.T., Ivison, R.J., Scoville, N.Z., Yun, M., Evans, A.S., Smail, I., Blain, A.W. \& Kneib, J.-P., 1998, \apjl, 505, L10
\bibitem[Garmire et al.(2000)]{ga00} Garmire, G.P. et al., 2000, in preparation.
\bibitem[Gunn \& Shanks(1999)]{gs99} Gunn, K. \& Shanks, T., astroph/990989
\bibitem[Hasinger(2000)]{ha00} Hasinger, G., astroph/0001360
\bibitem[Hughes et al.(1998)]{hu98} Hughes, D.H. et al., 1998, Nature, 394, 241
\bibitem[Holland et al.(1998)]{hol98} Holland, W. S. et al., 1999, \mnras, 303, 659
\bibitem[Hornschemeier et al.(2000)]{hor00} Hornschemeier et al., 2000, \apjl,
in press (astroph/0004260)
\bibitem[Ivison et al.(1998)]{iv98} Ivison, R. J. et al., 1998, \mnras, 298, 583 (I98)
\bibitem[Iwasawa et al.(1997)]{iw97} Iwasawa, K., Fabian, A.C. Matt, G., 1997, \mnras, 289, 443
\bibitem[Kneib et al.(1993)]{kn93} Kneib, J.-P., Mellier, Y., Fort, B. \& Mathez, G., 1993, \aap, 273, 367
\bibitem[Kraft, Burrows \& Nousek(1991)]{kbn91} Kraft, R. P., Burrows, D. N. \& Nousek, J. A., 1991, \apj, 374, 344
\bibitem[Mushotzky et al.(2000)]{mu00} Mushotzky, R., Cowie, L., Barger, A. \& Arnaud, K., 2000, Nature, 404, 459
\bibitem[Nandra et al.(1997)]{na97} Nandra, K., George, I. M., Mushotzky, R. F., Turner, T. J., \& Yaqoob, T., 1997, \apj, 477, 602
\bibitem[Ogasaka et al.(1997)]{og97} Ogasaka, Y., Inoue, H., Brandt, W. N., Fabian, A. C., Kii, T., Nakagawa, T., Fujimoto, R. \& Otani, C., 1997, \pasj, 49, 179
\bibitem[Petrucelli, Nandram \& Chen(1999)]{pe99} Petrucelli, J. D., Nandram, B. 
\& Chen, M., 1999, {\em Applied Statistics for Engineers and Scientists} (Saddle River, NJ: Prentice Hall) p 682.
\bibitem[Severgnini et al.(2000)]{se00} Severgnini, P., et al., 2000, \aap, in press; astroph/0006233
\bibitem[Smail et al.(1997)]{sm97} Smail, I., Ivison, R.J., Blain, A.W., 1997,
\apjl, 490, L5
\bibitem[Smail et al.(1998)]{sm98} Smail, I., Ivison, R.J., Blain, A.W., \&
Kneib, J.P., 1998, \apjl, 507, L21
\bibitem[Soucail et al.(1999)]{so99} Soucail, G., Kneib, J.P., Bezecourt, J., Metcalfe, L., Altieri, B., \& le Borgne, J.F., 1999, \aap, 343, L70 (S99)
\bibitem[Stark et al.(1992)]{st92} Stark, A. A., Gammie C.P., Wilson R.W., Bally J., Linke R., Heiles C. \& Hurwitz M., 1992, \apjs, 79, 77
\bibitem[Turner et al.(1997)]{tu97} Turner, T.J., George, I.M., Nandra, K. \& Mushotzky, R.F., 1997, \apjs, 113, 23
\bibitem[Vignali et al.(1999)]{vi99} Vignali, C., Comastri, A., Cappi, M., Palumbo, G.G.C., Matsuoka, M. \& Kubo, H., 1999, \apj, 516, 582
\bibitem[Vignati et al.(1999)]{vit99} Vignati, P., et al., 1999, \aap, 349, L57
\end{thebibliography}
\end{document}